\documentstyle[preprint,prb,aps,amsfonts]{revtex}

\input epsf

\newcommand{\cov}{\nabla}
\newcommand{\R}{\mbox{${\Bbb R}$}}
\newcommand{\M}{\mbox{${\cal M}$}}
\newcommand{\A}{\mbox{${\cal A}$}}
\newcommand{\spacetime}{\mbox{$(\M,g)$}}
\newcommand{\pdiff}[2]{\frac{\partial #2}{\partial #1}}
\newcommand{\Dd}[2]{\frac{D#2}{\partial #1}}
\newcommand{\E}{{E_0}}
\newcommand{\Ek}{{E_k}}
\newcommand{\ddt}{\pdiff t{}}
\newcommand{\ddu}{\pdiff u{}}
\newcommand{\Ddt}{\Dd t{}}
\newcommand{\Ddu}{\Dd u{}}
\newcommand{\Lie}{{\cal L}}
\newcommand{\proof}{\noindent{\it Proof.}\ }
\newcommand{\qed}{\hfill$\Box$}

\newtheorem{theorem}{Theorem}
\newtheorem{lemma}{Lemma}
\newtheorem{corollary}{Corollary}

\begin{document}

\tighten
\draft
\title{A Local Variational Theory for the Schmidt Metric}
\author{Fredrik St{\aa}hl\thanks{e-mail: {\tt fredriks@abel.math.umu.se}}}
\address{Department of Mathematics, University of Ume{\aa}, Ume{\aa}, 
Sweden}
\date{\today}

\maketitle
\begin{abstract}
We study local variations of causal curves in a space-time with 
respect to b-length (or generalised affine parameter length).  In a 
convex normal neighbourhood, causal curves of maximal metric length 
are geodesics.  Using variational arguments, we show that causal 
curves of minimal b-length in sufficiently small globally hyperbolic 
sets are geodesics.  As an application we obtain a generalisation of a 
theorem by B.\ G.\ Schmidt, showing that the cluster curve of a 
partially future imprisoned, future inextendible and future 
b-incomplete curve must be a null geodesic.  We give examples which 
illustrate that the cluster curve does not have to be closed or 
incomplete.  The theory of variations developed in this work provides 
a starting point for a Morse theory of b-length.
\end{abstract}

\narrowtext

\section{Introduction}\label{sec:intro}

The classical singularity theorems by Hawking and Penrose 
\cite{Hawking-Ellis,Penrose:Battelle} predict that in a physically 
reasonable model of space-time, singularities will inevitably occur in 
the form of incomplete, inextendible causal geodesics.  Unfortunately 
the nature of these singularities are still not fully understood.  
From a physical viewpoint one would expect some sort of divergence of 
the curvature when approaching a singularity.  More precisely, we 
might expect divergence of some scalar polynomial in the Riemann 
tensor and the metric along an incomplete curve ending 
at the singularity, or equivalently that the Riemann tensor components 
are unbounded in {\em all} frames along the curve.

There are however singular space-times where the scalar polynomials 
are bounded, i.e.\ there exists some frame where the Riemann tensor 
components are bounded, along an incomplete inextendible causal 
curve.  In this paper we concentrate on space-times containing 
b-incomplete causal curves partially imprisoned in a compact set.  The 
existence of imprisoned curves is equivalent to the existence of a 
cluster point to some causal curve, and thus implies a non-Hausdorff 
behaviour of the b-completion.  
\cite{Schmidt:b-boundary,Schmidt:b-complete,Hawking-Ellis}

A well-known example of totally imprisoned inextendible incomplete 
curves can be found in the Taub-NUT space-time.  Instead of analysing 
the Taub-NUT space-time directly we will consider a two-dimensional 
space-time with similar properties given by Misner.  
\cite{Misner:Taub-NUT} We will refer to this space-time as the Misner 
space-time.  The manifold is $S^1\times\R$ with the metric
\[ds^2=2\,dt\,d\psi+t\,d\psi^2\quad\text{for}\quad
	t\in\R\ \text{and}\ \psi\in[0,2\pi).\]
The vertical lines $t=u$, $\psi=const.$ are complete null geodesics.  
In addition there are families of null geodesics with affine parameter 
$u$ which follow infinite spirals as they approach $t=0$, of the type 
$t=Cu$, $\psi=-\ln u^2$ where $C$ is a non-zero constant (Fig.\ 
\ref{fig:misner}).  The affine parameter is bounded on these 
inextendible geodesics as they approach $t=0$ so they are in fact 
incomplete.  The closed null geodesic given by $t=0$, $\psi=-\ln u^2$ 
is incomplete, and there are no curvature singularities.

If we endow a neighbourhood of $t=0$ with a continuous frame field we 
find that the tangent vector to one of the spiralling incomplete 
curves are boosted by an unbounded amount relative to the frame field 
as the curve approaches $t=0$.  The reason that the Riemann components 
remain bounded along the curve is that the Riemann tensor is highly 
specialised in this case, and it is likely that the situation is 
unstable because of the back reaction of test particle travelling 
along the curve.\cite{Tipler-Clarke-Ellis}

Indeed one can show that a b-boundary point $p$ corresponding to an 
inextendible incomplete causal curve $\lambda$ must be a p.p.\ 
singularity, i.e.\ the Riemann tensor components diverge in any 
parallelly propagated frame along the curve, if either 
{\em i)} space-time is globally hyperbolic with the Riemann tensor 
non-specialised at $p$ 
\cite{Clarke:sing-glob-hyp,Clarke:singularities,Clarke:extensions2} or 
{\em ii)} $\lambda$ has a cluster point where the Riemann tensor is 
non-specialised.\cite{Hawking-Ellis}

Schmidt \cite{Schmidt:b-complete} has found the following result 
concerning totally imprisoned curves:

\begin{theorem}\label{thm:totally-imprisoned}
If a future inextendible, future b-incomplete causal 
curve is totally future imprisoned in a compact set $\cal N$, 
then there is a future inextendible null geodesic 
totally imprisoned in $\cal N$.
\end{theorem}

Our aim is now to develop a local theory of variations for b-length, and 
to use this to prove a similar theorem for partially imprisoned 
curves.  Note that partial imprisonment is a weaker concept than total 
imprisonment, since a totally imprisoned curve is also partially 
imprisoned in a small compact neighbourhood of one of its cluster 
points.

The plan of the paper is as follows.  In section \ref{sec:defs} we 
discuss the appropriate definitions and section \ref{sec:vari} 
contains the variational arguments.  The generalisation of Theorem 
\ref{thm:totally-imprisoned} is provided in section \ref{sec:impr} and 
section \ref{sec:conc} is devoted to discussions and examples.

\section{Definitions}\label{sec:defs}

Similar to the terminology in Ref.\ \onlinecite{Hawking-Ellis}, 
\spacetime\ is a space-time consisting of a connected four-dimensional 
Hausdorff $C^{\infty}$ manifold \M\ and a Lorentz metric $g$ of 
signature $\hbox{(-~+~+~+)}$.  Throughout the rest of this paper, 
``incomplete'' will mean ``future incomplete'', ``incomplete'' is 
``future incomplete'' and so on.  This is for convenience only, since 
all results hold equally well for past incomplete, past inextendible 
curves.

Let $\lambda(t)$ be a causal curve in \M\ from
$p=\lambda(t_p)$ to $q=\lambda(t_q)$ with tangent vector $V$.  
The {\em metric length} of $\lambda$ is given by
\begin{equation}\label{eqn:Ldef}
	L(\lambda):=\int_{t_p}^{t_q}\left[-g(V,V)\right]^{\frac12}\,dt.
\end{equation}
The generalised affine parameter length, or {\em bundle length} of 
$\lambda$ with respect to a given orthonormal basis 
$\Ek$ is
\begin{equation}\label{eqn:gapdef}
	l(\lambda,\Ek):=\int_{t_p}^{t_q}
	\left[\sum_{k=0}^3(V^k)^2\right]^{\frac12}\,dt
\end{equation}
where $V^k$ are the components of the tangent vector $V$ in the basis 
$\Ek$.\cite{Hawking-Ellis} Note that the b-length coincides with 
affine parameter length for geodesics and that for causal curves, the 
b-length is always greater than the metric length, i.e.  
\[l(\lambda,\Ek)\ge L(\lambda).\]

Let $\phi:\A=(-\epsilon,\epsilon)\times[t_p,t_q]\to\M$ be a 
one-parameter causal variation of $\lambda$ in \M, i.e.
$\phi$ satisfies the conditions
\begin{enumerate}
	\item $\phi(0,t)=\lambda(t)$

	\item $\phi$ is ${\cal C}^3$ except at a finite number of points

	\item $\phi(u,t_p)=p$ and $\phi(u,t_q)=q$ for all 
	$u\in(-\epsilon,\epsilon)$ 

	\item $\phi(u_0,t)$ is a causal curve for each 
	$u_0\in(-\epsilon,\epsilon)$.
\end{enumerate}
Starting with an orthonormal basis $\Ek$ at $p$ with $\E$ timelike, we 
parallelly propagate $\Ek$ along $\phi(u_0,t)$ for each constant 
$u_0$, obtaining functions $\A\to T_{\phi(u,t)}\M$, 
$(u,t)\mapsto\Ek(u,t)$ for $k=0,1,2,3$.  We denote the tangent vectors 
in $T_{(u,t)}\A$ with respect to $u$ and $t$ by $\ddu$ and $\ddt$ and 
their corresponding vectors in $T_{\phi(u,t)}\M$ by $X:=\phi_*\ddu$ 
and $V:=\phi_*\ddt$ respectively.

The map $\phi$ is not necessarily injective, which means that the 
vectors $V$, $X$ and $\E$ do not in general constitute vector fields 
in \M.  This could prove to be a problem when we try to study the 
variation of $l(\lambda,\Ek)$.  The problem can be avoided however if 
we {\em define} the covariant derivatives needed as 
\cite{Felice-Clarke}
\begin{mathletters}
\begin{equation}
	(\Ddt\E)^a:=\frac{d}{dt}\E^a+\Gamma^a_{bc}\E^bV^c
\end{equation}
\begin{equation}
	(\Ddu\E)^a:=\frac{d}{du}\E^a+\Gamma^a_{bc}\E^bX^c
\end{equation}
\begin{equation}
	(\cov_{\E}X)^a:={X^a}_{,b}\E^b+\Gamma^a_{bc}X^b\E^c.
\end{equation}
\end{mathletters}
It is then possible to do the usual calculations, in 
particular 
\begin{equation}\label{eqn:jacobi}
	\Ddt\Ddu\E=R(V,X)\E+\Ddu\Ddt\E+\cov_{[V,X]}\E=R(V,X)\E
\end{equation}
since $[V,X]=0$ and $\Ddt\E=0$.  

For $V,W\in T_{\phi(u,t)}\M$ let 
\begin{equation}\label{eqn:hdef}
	h_{(u,t)}(V,W):=g(V,W)+2g(V,\E(u,t))g(W,\E(u,t)).
\end{equation}
We may then rewrite (\ref{eqn:gapdef}) as 
\begin{equation}\label{eqn:ldef}
	l(\lambda,\Ek):=\left.\int_{t_p}^{t_q}
	\left[h_{(u,t)}(V,V)\right]^{\frac12}dt\right|_{u=0}.
\end{equation}
Again, $h_{(u,t)}$ might not be a tensor field since $\E(u,t)$ might 
have different values at the same point in \M.
From now on we will write $h$ for $h_{(u,t)}$ and $\E$ for $\E(u,t)$, 
the dependence of $u$ and $t$ being understood. We will refer to $h$ 
as the {\em Schmidt metric}.

Given two timelike separated points $p$ and $q$ in \M\ we denote the 
space of continuous causal curves from $p$ to $q$ by $C(p,q)$, and the 
subset of $C(p,q)$ consisting of timelike $C^1$ curves by $C'(p,q)$.  
On these spaces we will use the $C^0$ topology where a neighbourhood 
of $\gamma\in C(p,q)$ is defined by all curves in $C(p,q)$ which lie 
in a neighbourhood of the image of $\gamma$ in \M.\cite{Hawking-Ellis} 
Note that this topology is non-Hausdorff if strong causality is 
violated, since it does not distinguish how many times a closed curve 
is traversed.  For the discussion at hand this problem is avoided 
since we are only interested in the structure on some small globally 
hyperbolic neighbourhood, and such a set can always be found around 
any point.  In this context, $C'(p,q)$ is dense in $C(p,q)$.  
(Remember that a globally hyperbolic set is a strongly causal set 
$\cal U$ where $C(p,q)$ is compact for all points $p,q\in{\cal U}$.  
\cite{Hawking-Ellis})

\section{Variational Theory}\label{sec:vari}

In this section we develop a theory of variations for b-length by 
adapting the usual procedure for metric length.\cite{Hawking-Ellis} We 
will be concerned with the properties of curves in a small globally 
hyperbolic set, and we will also need to restrict the set further to 
put bounds on the additional terms introduced by the parallelly 
propagated basis used in the b-length definition.  We start by 
computing the first variation of (\ref{eqn:ldef}).
 
\begin{lemma}\label{lemma:firstvar}
The first variation of $l(\lambda,\Ek)$ is
\begin{equation}\label{eqn:firstvar}
	\frac d{du} l(\lambda,\Ek)=
	-\int_{t_p}^{t_q}f^{-1}h(\Ddt V,P^h_{V^{\perp}}X)dt
	 + 2\int_{t_p}^{t_q}f^{-1}g(V,\E)g(V,\Ddu\E)dt
\end{equation}
where $f:=\left[h(V,V)\right]^{\frac12}$ and
$P^h_{V^{\perp}}$ is the projection with respect to $h$ onto
the space of tangent vectors orthogonal to $V$.
\end{lemma}
\proof
First note that from (\ref{eqn:hdef}) and $\Lie_XV=0$ we have 
\begin{equation}\label{eqn:dduhVV}
	\ddu[h(V,V)]=4g(V,\E)g(V,\Ddu\E)+2h(\Ddt X,V).
\end{equation}
We use this to rewrite $\frac d{du} l(\lambda,\Ek)$ as 
\[\frac d{du} l(\lambda,\Ek)=
	\int_{t_p}^{t_q}\ddu
	\left(\left[h(V,V)\right]^{\frac12}\right)dt\]
\[=\int_{t_p}^{t_q}f^{-1}h(\Ddt X,V)\,dt
	 + 2\int_{t_p}^{t_q}f^{-1}g(V,\E)g(V,\Ddu\E)\,dt.\]
The first term can now be reformulated as a sum involving a total 
$t$ derivative which vanishes because of the fixed endpoints of 
the variation.
\[\int_{t_p}^{t_q}f^{-1}h(\Ddt X,V)\,dt
	=\int_{t_p}^{t_q}\ddt\left[f^{-1}h(X,V)\right]dt
	- \int_{t_p}^{t_q}f^{-1}h(\Ddt V,X-f^{-2}h(X,V)V)\,dt.\]
Defining the projection with respect to the Schmidt metric $h$ as 
$P^h_{V^{\perp}}=X-f^{-2}h(X,V)V$ we get (\ref{eqn:firstvar}).
\qed

The first term in Lemma \ref{lemma:firstvar} is similar to the 
expression occurring in the metric case (c.f.\ Ref.\ 
\onlinecite{Hawking-Ellis}).  The additional second term corresponds 
to the variation of the basis vector $\E$ when parallelly transporting 
$\E$ along different curves.  The change in $\E$ is determined by the 
Riemann tensor and one would expect it to be small for short curves.  
Indeed, we have

\begin{lemma}\label{lemma:bound}\ 
\begin{equation}\label{eqn:bound}
	\left.g(V,\E)g(V,\Ddu\E)\right|_t=
	K(t)\sup_{[t_p,t_q]}\left\{h(X,X)^{\frac12}\right\}
\end{equation}
where $K(t)=O(t-t_p)$.
\end{lemma}
\proof
We start by rewriting the second factor on the left side as
\[g(V,\Ddu\E)=g_{\hat{a}\hat{b}}
	V^{\hat{a}}\E^{\hat{b}}{}_{;\hat{c}}X^{\hat{c}}\]
where hatted indices denotes components with respect to the basis
$\Ek$.  Now let 
\[r:= \sup_{[t_p,t_q]}\left\{
	\left\|{R^{\hat{c}}}_{\hat{0}\hat{a}\hat{b}}
	A^{\hat{a}}B^{\hat{b}}\right\|;
	A,B\in\R^4,\|A\|=\|B\|=1\right\}\]
\[g:= \sup_{[t_p,t_q]}\left\{
	\left|g_{\hat{a}\hat{b}}A^{\hat{a}}B^{\hat{b}}\right|;
	A,B\in\R^4,\|A\|=\|B\|=1\right\}\]
where $\|\cdot\|$ is the usual Euclidian norm in $\R^4$.  Then
\[\left.\E^{\hat{b}}{}_{;\hat{c}}X^{\hat{c}}\right|_t=\int_{t_p}^t 
	(\E^{\hat{b}}{}_{;\hat{c}}X^{\hat{c}})_{;\hat{d}}V^{\hat{d}}\,dt
	=\int_{t_p}^t {R^{\hat{b}}}_{\hat{0}\hat{c}\hat{d}}
	X^{\hat{c}}V^{\hat{d}}\,dt\]
by (\ref{eqn:jacobi}), and
\[\left\|\int_{t_p}^t {R^{\hat{b}}}_{\hat{0}\hat{c}\hat{d}}
	X^{\hat{c}}V^{\hat{d}}\,dt\right\|
	\le r(t-t_p){\|V\|}_{\lambda}{\|X\|}_{\lambda}\]
where ${\|\cdot\|}_{\lambda}:=
\sup_{[t_p,t_q]}\left\{h(\cdot,\cdot)^{\frac12}\right\}$.
It follows immediately that
\[\left|g(V,\E)g(V,\Ddu\E)\right|\le 
	rg^2{\|V\|}_{\lambda}^3(t-t_p){\|X\|}_{\lambda}.\]
\qed

In the metric case, timelike geodesics are curves of extremal metric length 
provided there are no conjugate points along the geodesic.  In order 
to prove a similar result for the b-length, we need to construct 
a variation of any non-geodesic causal curve which results in a 
shorter curve.  This is done in the following somewhat lengthy lemma.

\begin{lemma}\label{lemma:b-shorter}
Let $\lambda$ be a non-geodesic causal curve from $p$ to $q$.  
Then there is a variation of $\lambda$ giving a causal curve from 
$p$ to $q$ with smaller b-length than $\lambda$.
\end{lemma}
\proof 
We want to construct a causal variation of $\lambda$ with 
variation vector $X$ such that $\ddu[h(V,V)]<0$.  We proceed by 
the following steps:
\begin{enumerate}
	\item Choose $X$ such that the second term in (\ref{eqn:dduhVV}) 
	is everywhere non-positive, and strictly negative somewhere, i.e.
	\begin{equation}\label{eqn:b-negative}
		h(\Ddt X,V)\le0
	\end{equation}
	with strict inequality on some part of $\lambda$.

	\item Check that the variation is causal.

	\item Show that by a suitably small variation along $X$ and by 
	restricting the variation to a sufficiently small portion of 
	$\lambda$, the second term in (\ref{eqn:dduhVV}) dominates the 
	first, i.e.
	\begin{equation}\label{eqn:b-estimate}
		g(V,\E)g(V,\Ddu\E)<\frac12\left|h(\Ddt X,V)\right|.
	\end{equation}
\end{enumerate}

\noindent{\it Case 1.} 
We start with the case when the tangent vector $V$ is continuous.  Let 
$\lambda$ be parameterised by b-length.  Then $h(V,V)=1$ and 
\begin{equation}\label{eqn:orthogonal}
	0=\ddt[h(V,V)]=2h(V,\Ddt V)
\end{equation}
so $\Ddt V$ is orthogonal to $V$ with respect to $h$ whenever $\Ddt V$ 
is non-zero, which has to be the case somewhere along $\lambda$ since 
$\lambda$ is non-geodesic.

Let the variation vector be 
\begin{equation}\label{eqn:Xdef}
	X:=xV+y\Ddt V
\end{equation}
where $x$ and $y$ are functions vanishing outside some interval 
$[\alpha,\beta]\subset[t_p,t_q]$ such that
$\Ddt V\neq0$ on $[\alpha,\beta]$.  

\noindent{\em Step 1.} We restrict our attention to the interval
$[\alpha,\beta]$.  On this interval, (\ref{eqn:orthogonal}) and 
(\ref{eqn:Xdef}) implies 
\[h(\Ddt X,V)=\ddt\left[h(X,V)\right]-h(X,\Ddt V)
	=\frac{dx}{dt}-ay\]
where $a:=h(\Ddt V,\Ddt V)$.  Choose $x$ as
\[x(t):=\int_{\alpha}^t(ay-1)\,dt.\]
Then 
\[h(\Ddt X,V)=-1\] 
for any function $y$ on $[\alpha,\beta]$ with
\[y(\alpha)=y(\beta)=\int_{\alpha}^{\beta}(ay-1)\,dt=0.\]
We may choose $y$ to be positive.

\noindent{\em Step 2.} If $\lambda$ has a timelike segment, a 
sufficiently small variation of that segment will give a timelike 
curve.  If $\lambda$ is null, $g(V,V)=0$ gives
\begin{equation}\label{eqn:hVV=1}
	1 = h(V,V) = 2g(V,\E)^2
\end{equation}
so $g(V,\E)^2=\frac12$. Using $\Lie_XV=0$ and the definition of $h$ 
(Eq.\ \ref{eqn:hdef}) we then get 
\[\ddu\left[g(V,V)\right]
	=2\ddt\left[g(X,V)\right] - 2g(X,\Ddt V)\]
\[=2\ddt\left[h(X,V)\right] - 2h(X,\Ddt V)
	 - 4\left(\ddt\left[g(X,\E)g(V,\E)\right]
	 - g(X,\E)g(\Ddt V,\E)\right)\]
\[=-2 - 4\ddt\left[g(X,\E)\right]g(V,\E)\]
where we have used $h(\Ddt X,V)=-1$ to obtain the last equality. 
Finally, (\ref{eqn:hVV=1}) implies $g(\Ddt V,\E)=0$ so 
\[\ddu\left[g(V,V)\right]
	=-2 - 4\frac{dx}{dt}g(V,\E)^2
	=-2ay<0\]
since both $a$ and $y$ are positive on $[\alpha,\beta]$, and hence the 
curve will remain causal under a small variation along $X$.

\noindent{\em Step 3.} We restrict $y$ such that $ay\le2$ and 
$[\alpha,\beta]$ such that $\beta-\alpha\le1$ and 
$K<\frac16\inf_{[\alpha,\beta]}\{1,\sqrt{a}\}$ in 
(\ref{eqn:bound}) applied to $\lambda|_{[\alpha,\beta]}$.  Then Lemma 
\ref{lemma:bound} gives 
\[g(V,\E)g(V,\Ddu\E)\le
	K\sup_{[\alpha,\beta]}\left\{h(X,X)^{\frac12}\right\}\]
\[\le K\sup_{[\alpha,\beta]}\left\{|x|+\sqrt{a}y\right\}
	\le K\sup_{[\alpha,\beta]}\{|ay-1|\}(\beta-\alpha)
	 + K\sup_{[\alpha,\beta]}\{\sqrt{a}y\}\]
\[<\frac16+\frac16\sup_{[\alpha,\beta]}\{ay\}\le\frac12.\]
Thus $\ddu\left[h(V,V)\right]<0$ and the lemma is true for 
curves with a continuous tangent vector.

\noindent{\it Case 2.} 
Suppose now that $\lambda$ is made up of a finite number of geodesic 
segments.  Let $\lambda$ be parameterised by b-length which is 
equivalent to affine parameterisations of the geodesic segments.  It 
is sufficient to study the case when $V$ is discontinuous at one point 
$\lambda(t_0)$.  Let $W$ be the discontinuity at $\lambda(t_0)$, i.e.
\[W:=V_+-V_-\]
where
\[V_+:=\lim_{t\to t_0^+}V\qquad\text{and}\qquad
	V_-:=\lim_{t\to t_0^-}V_{\lambda(t)}.\]
Parallelly propagate $W$ along $\lambda$.  Then
\[\ddt\left[h(W,V)\right]=h(\Ddt W,V)+h(W,\Ddt V)=0\]
on each geodesic section of $\lambda$.  We know that 
\[\left.h(W,V)\right|_{\lambda(t)}=
	\left.h(W,V_-)\right|_{\lambda(t_0)}=
	h(V_+,V_-)-1\]
if $t\in[t_p,t_0)$ and
\[\left.h(W,V)\right|_{\lambda(t)}=
	\left.h(W,V_+)\right|_{\lambda(t_0)}=
	1-h(V_+,V_-)\]
if $t\in(t_0,t_q]$, i.e.\ $h(W,V)$ is negative on $[t_p,t_0)$ and 
positive on $(t_0,t_q]$.  Let the variation vector be $X:=xW$ where
\begin{equation}\label{eqn:xdef}
	x(t):=\left\{\begin{array}{ll}
	-h(W,V)^{-1}(\beta-t_0)(t-\alpha)&\text{when $t\in[\alpha,t_0)$}\\
	h(W,V)^{-1}(\beta-t)(t_0-\alpha)&\text{when $t\in(t_0,\beta]$}
	\end{array}\right.
\end{equation}
on some interval $[\alpha,\beta]\subset[t_p,t_q]$ to be chosen below, 
and zero otherwise.  

\noindent{\em Step 1.} On the interval $[\alpha,\beta]$,
\[h(\Ddt X,V)=
	\frac d{dt}[xh(W,V)]=
	\left\{\begin{array}{ll}
		-(\beta-t_0)&\text{when $t\in[\alpha,t_0)$}\\
		-(t_0-\alpha)&\text{when $t\in(t_0,\beta]$}
	\end{array}\right.\]
which is negative.

\noindent{\em Step 2.}
If one of the geodesic segments is null, we must ensure that the 
varied curve remains causal.  Since $\Lie_XV$, $\Ddt V$ and $\Ddt W$ 
all vanish we have 
\[\ddu\left[g(V,V)\right]=
	\frac d{dt}\left[xg(W,V)\right]=
	g(W,V)\frac {dx}{dt}.\]
Suppose that $V_-$ is null. On $[\alpha,t_0]$,
\[g(W,V)=g(V_+,V_-)-g(V_-,V_-)=g(V_+,V_-)\le0\] 
since $V_+$ and $V_-$ are causal vectors. But on this interval 
\[\frac{dx}{dt}=-h(W,V)^{-1}(\beta-t_0)\ge0\]
so $\ddu\left[g(V,V)\right]\le0$. The case when $V_+$ is null is 
similar.

\noindent{\em Step 3.}
We choose $[\alpha,\beta]$ such that $\beta-\alpha\le1$ and
\[K<\frac1{2\sqrt2}\left(1-h(V_+,V_-)\right)^{\frac12}\]
in (\ref{eqn:bound}). Lemma \ref{lemma:bound} gives 
\begin{equation}\label{eqn:estimate}
	g(V,\E)g(V,\Ddu\E)\le
	K\sup_{[\alpha,\beta]}\left\{h(X,X)^{\frac12}\right\}
	=K\sup_{[\alpha,\beta]}\left\{
	|x|\left(2-2h(V_+,V_-)\right)^{\frac12}\right\}.
\end{equation}
Now we can use the definition of $x$ (Eq.\ \ref{eqn:xdef}) to get an estimate.
\[|x|\le|h(V,W)|^{-1}(\beta-t_0)(t_0-\alpha)\]
\[\le K\sup_{[\alpha,\beta]}\left\{
		\left(1-h(V_+,V_-)\right)^{-1}
		\left(2-2h(V_+,V_-)\right)^{\frac12}
		\right\}(\beta-t_0)(t_0-\alpha).\]
Substituting this into (\ref{eqn:estimate}) gives 
\[g(V,\E)g(V,\Ddu\E)<\frac12(\beta-t_0)(t_0-\alpha)\]
so again $\ddu\left[h(V,V)\right]<0$.
\qed

We have now established that if a causal curve has minimal b-length, 
it must be a geodesic.  It remains to prove the existence of causal 
curves with minimal b-length.  First we need a result on the 
continuity properties of the b-length.

\begin{lemma}\label{lemma:semicont}
Suppose that all the curves in $C'(p,q)$ are contained in a strongly causal 
region.
Then the b-length $l$ is lower semi-continuous in the 
$C^0$-topology on $C'(p,q)$.	
\end{lemma}
\proof
Let $\lambda\in C'(p,q)$ be a timelike curve from $p$ to $q$, 
parameterised by b-length $t$ such that the map $t\mapsto\lambda(t)$ 
is injective.  This has to be possible since otherwise strong 
causality would be violated.  Let $f$ be a function in a neighbourhood 
${\cal U}$ of $\lambda$ such that $f|_{\lambda}=t$ and the surfaces of 
constant $f$ are spacelike and orthogonal to the tangent $V$ of 
$\lambda$ with respect to $h$.  Any curve $\mu\in C'(p,q)\cap{\cal U}$ 
can be parameterised by $f$, and the tangent vector of $\mu$ may be 
expressed as
\[\left.\pdiff f{}\right|_{\mu}= Z+W,\]
where $Z^a=g^{ab}f_{;b}$ and $h(Z,W)=0$.  Then
\[h\left(\left.\pdiff f{}\right|_{\mu},\left.\pdiff f{}\right|_{\mu}\right)=
	h(Z,Z)+h(W,W) \ge h(Z,Z).\]
But $Z|_{\lambda}=V$ and $\lambda$ is parameterised by b-length, so 
$h(Z|_{\lambda},Z|_{\lambda})=1$.  Given $\epsilon>0$ we can choose a 
neighbourhood ${\cal U}'\subset{\cal U}$ of $\lambda$ such that 
$h(Z,Z)>1-\epsilon$ on ${\cal U}'$.  Then for all curves $\mu$ in 
${\cal U}'$,
\[l(\mu,\Ek)\ge\int_{t_p}^{t_q}h(Z|_{\mu},Z|_{\mu})^{\frac12}\,dt>
	\sqrt{1-\epsilon}\int_{t_p}^{t_q}dt=
	\sqrt{1-\epsilon}\ l(\lambda,\Ek).\]
\qed

We summarise the results of this section in a theorem.

\begin{theorem}\label{thm:geodesic}
If $p$ and $q$ are causally separated and belong to a globally 
hyperbolic set, there exists a causal curve from $p$ to $q$ with 
minimal b-length.  Moreover, any such curve is geodesic.
\end{theorem}
\proof
$l$ is lower semi-continuous by Lemma \ref{lemma:semicont} and bounded 
below by $0$ on the closure of $C'(p,q)$, which is compact since $p$ 
and $q$ belong to a globally hyperbolic set.  Thus there is a curve 
$\gamma$ in the closure of $C'(p,q)$ with minimal b-length.  $\gamma$ 
must be geodesic since if otherwise, $\gamma$ can be varied to give a 
shorter curve by Lemma \ref{lemma:b-shorter}.
\qed

\section{Imprisoned curves}\label{sec:impr}

We can now use Theorem \ref{thm:geodesic} to generalise 
Schmidt's theorem (Theorem \ref{thm:totally-imprisoned}) to partially 
imprisoned curves.

\begin{theorem}\label{thm:cluster}
Every cluster curve $\gamma$ of a partially imprisoned incomplete 
inextendible causal curve $\lambda$ is a null geodesic.
\end{theorem}

\proof
The intuitive picture is that the tangent of $\lambda$ must become 
more and more null as one follows $\lambda$ to the future.  We will 
prove this by contradiction;\ if $\gamma$ is timelike somewhere, the 
b-length of $\lambda$ must be infinite.

If $\gamma$ is not a null geodesic, we can find a small portion 
$\gamma'$ of $\gamma$ contained in a globally hyperbolic, convex 
normal neighbourhood ${\cal N}$ such that the endpoints $p$ and $q$ of 
$\gamma'$ are timelike separated.  We can also find portions 
$\lambda_i$ of $\lambda$ contained in ${\cal N}$ such that $\gamma'$ 
is the limit curve of the $\lambda_i$.
		
Let $\pi$ be the b-shortest causal curve between $p$ and $q$ and let 
$\pi_i$ be the b-shortest causal curve between the endpoints of 
$\lambda_i$.  Then $\pi$ and $\pi_i$ are geodesics by Theorem 
\ref{thm:geodesic} and $l(\lambda_i)\ge l(\pi_i)\ge L(\pi_i)$.  In a 
convex normal neighbourhood, geodesics are uniquely defined as curves 
with maximal metric length between two points, so $L(\pi_i)\to L(\pi)$ 
as $i\to\infty$.

Now $l(\lambda)\ge\sum_i l(\lambda_i)$ and $l(\lambda)$ is finite 
since $\lambda$ is incomplete.  But $p$ and $q$ are timelike separated 
so 
\[l(\lambda_i)\ge L(\pi_i)\to L(\pi)>0 
\qquad\text{as}\qquad i\to\infty\] 
which is a contradiction.
\qed

\begin{figure}[tbp]
\centerline{\epsfbox{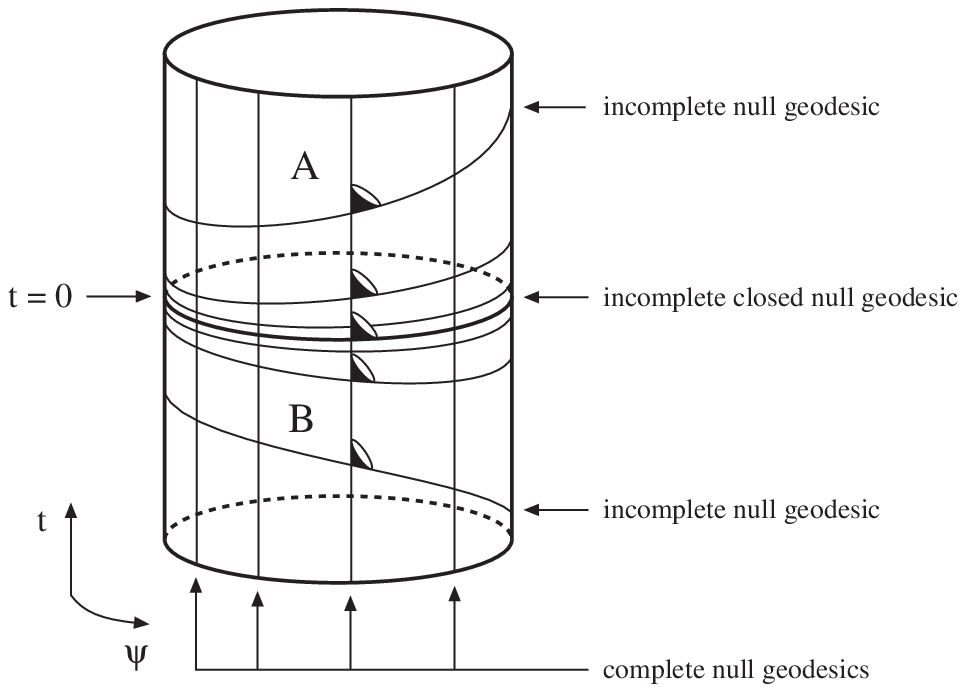}\vspace{1cm}}
\caption{Misner's two-dimensional space-time.  In region A causality 
is preserved but in region B there are closed timelike curves through 
every point.  The horizon is generated by the closed incomplete null 
geodesic at $t=0$.}
\label{fig:misner}
\end{figure}

In the Misner space-time (c.f.\ section \ref{sec:intro} and Fig.\ 
\ref{fig:misner}) the statement of Theorem \ref{thm:cluster} is not 
surprising since the closed null geodesic at $t=0$ is a cluster curve 
of totally imprisoned null geodesics.  In this particular case the 
cluster curve is inextendible and incomplete, but that is not always 
true as the following example shows.

\noindent{\it Example.}
Consider the manifold $S^1\times\R$ with the metric
\[ds^2=2dtd\psi+t^2d\psi^2.\]

\begin{figure}[tbp]
\centerline{\epsfbox{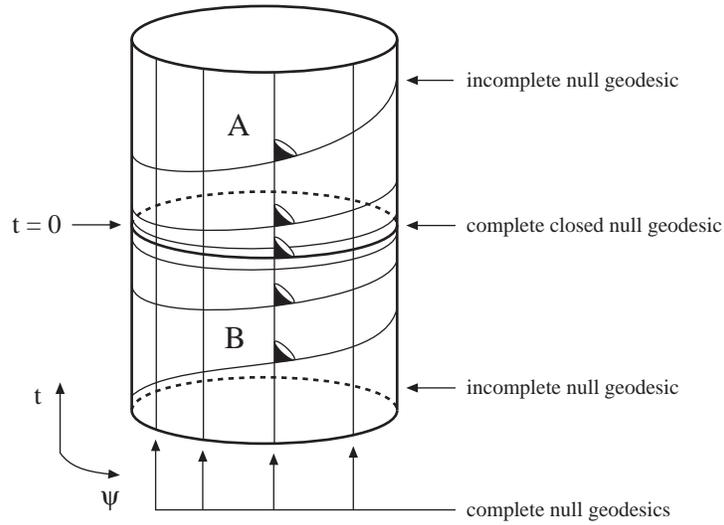}\vspace{1cm}}
\caption{A two-dimensional space-time with a complete cluster curve at 
$t=0$.  Causality is preserved in both regions A and B, but violated 
by the closed complete null geodesic at $t=0$.}
\label{fig:misnercomplete}
\end{figure}

This space-time exhibits a similar behaviour as the Misner space-time.  
There are complete vertical null geodesics given by $t=u$, 
$\psi=const.$ and incomplete, inextendible spiralling geodesics 
approaching the cluster curve at $t=0$ of the type $t=u$, 
$\psi=2u^{-1}$.  However, in this case the closed null geodesic is 
complete;\ it is given by $t=0$, $\psi=u$ (Fig.\ 
\ref{fig:misnercomplete}).

Another property of the cluster curve in the Misner space-time is that 
it is imprisoned in a compact set.  An example where this is not the 
case can be found by simply removing a point on $t=0$ in the Misner 
space-time (Fig.\ \ref{fig:misner}).

Note that closed timelike curves are partially imprisoned and that 
closed non-geodesic causal curves can be varied to give closed 
timelike curves.\cite{Hawking-Ellis}  Thus Theorem \ref{thm:cluster} 
implies the following.
\begin{corollary}
If a closed causal curve is non-geodesic or timelike, it must be 
complete.
\end{corollary}
This result is intuitively clear since a closed timelike curve 
$\lambda$ must be metrically complete.  When parallelly propagating an 
orthogonal basis along $\lambda$ we get an infinite sequence of 
orthogonal bases along the image of $\lambda$, each corresponding to 
one ``circulation'' of $\lambda$.  In each of these bases, $h(V,V)\ge 
g(V,V)$ where $V$ is the tangent of $\lambda$, with equality if and 
only if $V$ is orthogonal to the timelike basis vector $\E$ with 
respect to $g$.  Then the b-length has to be greater than or equal to 
the metric length which is infinite, so $\lambda$ must be complete.

\section{Concluding Remarks}\label{sec:conc}

The variational arguments used in this article are of a very local 
nature, and we have focused on the extremal properties of geodesics in 
small neighbourhoods.  The techniques are based on variations in a 
globally hyperbolic neighbourhood small enough to constrain the second 
term in the first variation of the b-length (\ref{eqn:firstvar}).  The 
next logical step is to shift attention to the global connection 
between geodesic behaviour and the b-length, i.e.\ to construct a 
Morse theory\cite{Milnor} for the Schmidt metric similar to the Morse 
theory for metric 
length.\cite{Uhlenbeck,Perlick:infinite,Perlick:Fermat,Giannoni-Masiello} 
The relation of conjugate points to extrema of the b-length is of 
particular interest.  This calls for some modifications of the 
techniques used in this article.  For example, causality might not 
hold in \M, which has the consequence that the topology on $C(p,q)$ is 
no longer Hausdorff.  A bigger problem is that in general, we expect 
that the curvature will have a severe impact on variations on a larger 
scale, since the parallel propagation of the basis used in the 
definition of b-length might give a large difference in b-length even 
for small variations.  This is subject to further study.

\section*{Acknowledgment}
I wish to express my gratitude towards my supervisor, Clarissa-Marie 
Claudel, for her invaluable support, comments and criticisms.

\end{document}